**SWIMSS: An Open Multi-Center SwinUNETR Model for Multiple Sclerosis White Matter Lesion Segmentation**

Federico Spagnolo[1,2,3,✝], Po-Jui Lu[1,2,3,✝], Mario Ocampo-Pineda[1,2,3], Lester Melie-Garcia[1,2,3], Esther Ruberte[1,2,3], Riccardo Galbusera[1,2,3,4], Matthias Weigel[1,2,3,5], Alessandro Cagol[1,2,3], Johanna Maria Lieb[6], Nicolin Hainc[7], Franca Wagner[8,9], Piotr Radojewski[10,11], Felix Tobias Kurz[12], Vincent Dunet[13], Emanuele Pravatà[14], Olaf Chan-Hi Kim[15], Patrick Roth[16], Marina Herwerth[16], Sebastian Finkener[17], Robert Hoepner[18], Patrice H. Lalive[19,20,21], Caroline Pot[22], Claudio Gobbi[23,24], Stefanie Müller[25], Jens Kuhle[2,3], Vincent Andrearczyk[26], Adrien Depeursinge[26,27], Cristina Granziera[1,2,3], for the Swiss MS Cohort

1. *Translational Imaging in Neurology (ThINk) Basel, Department of Biomedical Engineering, Faculty of Medicine, University Hospital Basel and University of Basel, Basel, Switzerland*
2. *Department of Neurology, University Hospital Basel, Basel, Switzerland*
3. *Research Center for Clinical Neuroimmunology and Neuroscience Basel (RC2NB), University Hospital Basel and University of Basel, Basel, Switzerland*
4. *Department of Neuropathology, University Medical Center Göttingen, Göttingen, Germany*
5. *Division of Radiological Physics, Department of Radiology, University Hospital Basel, Switzerland*
6. *Diagnostic and Interventional Neuroradiology, Clinic of Radiology and Nuclear Medicine, University Hospital Basel, Basel, Switzerland*
7. *Department of Neuroradiology, Clinical Neuroscience Center, University Hospital Zurich, University of Zurich, Zurich, Switzerland*
8. *Department of Diagnostic and Interventional Neuroradiology, Cantonal Hospital Aarau, Aarau, Switzerland*
9. *Department of Otorhinolaryngology and Head and Neck Surgery, University Hospital Bern and Faculty of Medicine, Bern, Switzerland*
10. *University Institute of Diagnostic and Interventional Neuroradiology, Inselspital, Bern University Hospital, University of Bern, Bern, Switzerland*
11. *Translational Imaging Center (TIC), Swiss Institute for Translational and Entrepreneurial Medicine, Bern, Switzerland*
12. *Division of Neuroradiology, Geneva University Hospitals, Geneva, Switzerland*
13. *Department of Diagnostic and Interventional Radiology, Lausanne University Hospital and University of Lausanne, Switzerland*
14. *Department of Neuroradiology, Neurocenter of Southern Switzerland, Lugano, Switzerland*
15. *Department of Radiology and Nuclear Medicine, H-OCH Health Ostschweiz & St. Gallen Medical School, Cantonal Hospital of St. Gallen, St. Gallen, Switzerland*
16. *Department of Neurology and Clinical Neuroscience Center, University Hospital Zurich and University of Zurich, Zurich, Switzerland*

17. *Department of Neurology, Kantonsspital Aarau, Aarau, Switzerland*
18. *Department of Neurology, Inselspital, Bern University Hospital, University of Bern, Bern, Switzerland*
19. *Department of Clinical Neurosciences, Division of Neurology, and Department of Medicine, Translational Biomarker Group, Geneva, Switzerland*
20. *Diagnostic Department, Division of Laboratory Medicine, Geneva University Hospital, Geneva, Switzerland*
21. *Department of Pathology and Immunology, Faculty of Medicine, University of Geneva, Geneva, Switzerland*
22. *Service of Neurology and Laboratories of Neuroimmunology, Department of Clinical Neurosciences, Lausanne University Hospital (CHUV) and University of Lausanne, Lausanne, Switzerland*
23. *Multiple Sclerosis Center, Neurocenter of Southern Switzerland, Ente Ospedaliero Cantonale, Lugano, Switzerland*
24. *Faculty of Biomedical Sciences, Università Della Svizzera Italiana (USI), Lugano, Switzerland*
25. *Department of Neurology, Cantonal Hospital St.Gallen, St. Gallen, Switzerland*
26. *Institute of Informatics, School of Management, HES-SO Valais-Wallis University of Applied Sciences and Arts Western Switzerland, Sierre, Switzerland*
27. *Nuclear Medicine and Molecular Imaging Department, Lausanne University Hospital (CHUV), Lausanne, Switzerland*

[‡]*these authors contributed equally to this work*

## Abstract

### Purpose

To develop and validate SWIMSS (SWInUNETR for Multiple Sclerosis white matter lesion Segmentation), a SwinUNETR-based deep learning (DL) model for reliable automated white matter (WM) lesion segmentation, in a large multi-centric cohort of people with multiple sclerosis (pwMS).

### Methods

A real-world longitudinal dataset of 1,385 pwMS from seven Swiss clinical centers was used to train, validate, and test SWIMSS and two state-of-the-art DL architectures: UNETR and MSSegDiff. Inference results obtained with LST-AI were also compared. Preprocessed magnetization-prepared rapid acquisition gradient echo and fluid-attenuated inversion recovery MRI sequences, acquired with various MR systems and field strengths, served as model inputs. Model performance was evaluated using the Dice similarity coefficient (DSC) with 95% confidence intervals (CI).

### Results

SWIMSS achieved the highest performance, with a DSC of 81.37 (95% CI: 80.63–82.11), outperforming UNETR (79.06; 95% CI: 78.33–79.79), MSSegDiff (75.70; 95% CI: 74.97–76.43), and LST-AI (63.19; 95% CI: 62.08–64.30). Normalized DSC followed a similar trend, with SWIMSS reaching 84.88. Lesion detection F1 achieved 81.14. Statistical analyses confirmed its superior performance ($p<0.001$).

### Conclusion

The SWIMSS achieved state-of-the-art performance for WM lesion segmentation. By making this model publicly available, we enable broader access to high-quality segmentation tools, allowing smaller centers—often lacking large annotated datasets or computational resources—to benefit from a model trained on a diverse, large-scale cohort. This promotes more equitable access to advanced imaging analysis and supports reproducibility and standardization in MS research.

## Introduction

Multiple sclerosis (MS) is a chronic immune-mediated neurological disease characterized by focal lesions visible on MRI. Lesion topography is central to current diagnostic criteria and involves specific regions of the white matter (WM), cortex, and spinal cord (1). Accurate WM lesion segmentation is essential, particularly in research settings, but manual delineation is time-consuming, requires expert raters, and is affected by inter-rater variability. Consequently, automated segmentation methods have emerged, although clinical adoption remains limited due to insufficient validation and poor workflow integration (2,3).

For automated WM lesion segmentation, Guarnera et al. (4) recently evaluated the performance of multiple networks on a 75-MS-patient dataset, encompassing architectures such as UNETR (5), SwinUNETR (6), and MSSegDiff — a denoising diffusion model proposed in Rondinella et al. (7) and then extended in Rondinella et al. (8) using the 15-patient ISBI 2015 dataset (9). UNETR and SwinUNETR achieved mean 5-fold Dice scores (DSC) of 64% and 68%, respectively, while MSSegDiff reached 69%. Dereskewicz et al. (10) combined public and private datasets, encompassing a total of 575 people with MS (pwMS), to train an nnU-Net model. The model achieved a DSC of 72% on the test set and a normalized DSC of 75%. LST-AI (11) achieved a DSC of 67% on smaller public datasets, although its deployment relied on a fixed inference framework, limiting adaptation to center-specific domain shifts. In contrast, datasets in other neurological domains often comprise thousands of patients (12,13). Building large multicenter cohorts is hindered by the need for harmonized annotations, data protection regulations, and intellectual property constraints (14).

The Swiss MS Cohort (SMSC) addresses some of these limitations by collecting clinical and imaging data from more than 2,200 patients across eight Swiss centers (15). A subset of head MRI scans with skull from 687 pwMS was used by Spagnolo et al. (16), to introduce an instance-level explanation method on 3D U-Net, nnU-Net, and SwinUNETR, focusing on lesion detection through a blob loss function. Among the three architectures, SwinUNETR achieved the best performance, with an F1 score of 73% and a DSC of 66%. In this work, we extend the state of the art by using a larger seven-center cohort, improving the automated segmentation pipeline, and releasing a publicly available model that supports both direct inference and fine-tuning. We further showed few-shot fine-tuning performances for pediatric MS to highlight the model's ability to adapt to in-house data.

## Materials and Methods

## Participants and Imaging Protocol

This study retrospectively used data from the SMSC (NCT02433028), approved by the local ethics review committee (Ethics Committee Northwest and Central Switzerland, BASEC-ID:023-02367). All participants provided written informed consent prior to participation. The inclusion criteria were: (1) age ≥ 18 years; (2) ability to provide informed consent; (3) MS diagnosis according to the 2017 McDonald criteria (17), or of a related neuroinflammatory condition (radiologically isolated syndrome [RIS] or neuromyelitis optica [NMO]) as defined in the cohort protocol (15); and (4) availability of a 3D brain MRI scan acquired by the end of 2024. Exclusion criteria encompassed major psychiatric or neurological comorbidities, contraindications to MRI, and pregnancy. Fluid-Attenuated Inversion Recovery (FLAIR) and Magnetization-Prepared Rapid Gradient-Echo (MPRAGE) images were acquired across 7 clinical centers with in-plane resolutions ranging from 0.5x0.5 to 1.3x1.3 $mm^2$ and slice thickness ranging from 0.5 to 1.3 *mm*.

Two public external datasets were also used to assess the out-of-domain generalizability of our approach: 1) MSSEG 2016 (18) 15-patient training set (as the test set has no publicly available ground truth masks, following Raab et al. (19)), with FLAIR and T1-weighted images of various resolutions ranging from 0.7x0.7x0.7 $mm^3$ to 1.2x1.0x1.0 $mm^3$ from three different centers; 2) pediMS (20), with longitudinal FLAIR and MPRAGE scans (0.75x0.75x3.9 $mm^3$ and 0.98x0.98x1.0 $mm^3$, respectively) from 9 pediatric participants with MS. This second dataset was also used to test the few-shot learning ability of the model (25 epochs, learning rate 5e-5).

### Data Preparation

The image preprocessing workflow for all datasets included skull-stripping using HD-BET (21), bias field correction and rigid co-registration using ANTs (22).

For the SMSC dataset, WM lesion masks were automatically generated from FLAIR and MPRAGE images (23) and manually corrected by 5 raters with >10 years experience in neuroimaging. The dataset was split into training (70%), validation (10%), and test (20%) subsets. To ensure balanced distributions across subsets, a stratified splitting strategy was employed based on both individual patients (to prevent data leakage) and total lesion volume.

### Data Augmentation and Sampling Strategy

Prior to model training, all images were resampled to 1 *mm* isotropic resolution. The images were cropped to the brain, maintaining a padding of 8 additional voxels in all directions to minimize unnecessary background processing and required GPU RAM. Voxel intensities within the brain were normalized using z-score normalization.

To address spatial class imbalance during training, patches were sampled dynamically using a targeted random cropping method based on positive and negative label ratios, with a target patch size of 96×96×96 voxels. This sampling strategy enforced a positive-to-negative label ratio of 4:1, extracting 5 patches per sample. On-the-fly stochastic data augmentation was applied during training to enhance model generalizability, including:

- Random 90-degree rotations with a probability of 0.5.
- Random intensity shifts with an offset of 0.3 and a probability of 0.5.
- Random affine transformations, introducing up to 15 degrees of rotation with respect to the axial direction and up to 10% spatial scaling.

### Model Architecture and Training Protocol

The SwinUNETR architecture configured with an initial embedding feature size of 24 elements (SWIMSS24) was trained for a total of 500 epochs and optimized using a combined Dice and Focal loss. The AdamW optimizer was utilized with a weight decay of 1e-3. The learning rate schedule incorporated a 32-epoch warm-up phase rising to 1e-3, after which it was adjusted to 3e-4 for the remaining training epochs. Post-processing retained only lesions larger than 3 voxels.

To evaluate the performance and robustness of the proposed pipeline, several comparator models were implemented for benchmarking:

- A SwinUNETR with the 12-element feature (SWIMSS12)
- A standard UNETR
- MSSegDiff, a generative diffusion-based segmentation network (8)
- LST-AI (11)

### Statistical Analysis

The primary metric for evaluating segmentation performance was the DSC. Normalized Dice (nDSC) and detection F1 score were computed as secondary outcome measures to provide a comprehensive assessment. Differences in DSC between the evaluated models were assessed using two-tailed paired Wilcoxon signed-rank tests. To account for multiple comparisons, *p*-values were adjusted using the Benjamini–Hochberg procedure. To quantify uncertainty, 95% confidence intervals for the paired DSC differences were estimated using 10,000 bootstrap iterations.

## Results

A total of 1,385 patients (66% females) and their longitudinal 4,912 1.5 or 3T 3D MRI acquisitions were sourced., with a median age at baseline (BL) of 41.1 years (IQR: 32.2 - 49.9;  range: 18.0 - 77.7). The prevalent disease course recorded at BL was relapsing MS (RMS, n=1226), followed by primary progressive MS (PPMS, n=73), secondary progressive MS (n=61), NMO (n=17), and RIS (n=8). The median Expanded Disability Status Scale score at BL was 2.0 (range: 0 - 8.5). The dataset also presented a median WM lesion volume of 4,341 $mm^3$ (range: 1,515, 12,525) $mm^3$.

The performance of the different architectures on the unseen test set is summarized in Table 1, where brackets on the top row specify the input patch size. SWIMSS24 achieved the best mean DSC of 81.37% (95% CI: 80.63-82.11), nDSC of 84.88% (95% CI: 84.19-85.56), and F1 score of 81.14% (95% CI: 80.44-81.85), followed by SWIMSS12(DSC: 80.84, 95% CI:80.11-81.57; nDSC: 84.36, 95% CI: 83.66-85.06; F1: 80.12, 95% CI: 79.41-80.82).

Table 2 summarizes the mean paired differences (model A - model B) and their 95% CI estimated with bootstrapping, with SWIMSS24 as the reference. The mean DSC of SWIMSS24 was statistically different from other models.

Figure 1 and 2 show, respectively, the predicted WM lesion masks generated by the best configurations of the four different networks (SWIMSS, UNETR, MSSegDiff, and LST-AI) for the test case with the highest lesion load and the test case with the highest number of WM lesions.

We also reported the inference execution time using GPU, peak system RAM and peak GPU RAM: SWIMSS24 [11.1 *s*, 1777 *MB*, 5970 *MB*], SWIMSS12 [9.0 *s*, 1627 *MB*, 2050 *MB*], UNETR [20.7 *s*, 2663 *MB*, 1888 *MB*], MSSegDiff [21.8 *s*, 2895*MB*, 10318*MB*], LST-AI [2 *min* 39 *s*, 2717 *MB*, 22596 *MB*]. Additionally, we computed execution time and peak system RAM when the inference is run on CPU only: SWIMSS24 [1 *min* 48 *s*, 2384 *MB*], SWIMSS12 [1 *min* 24 *s*, 2086*MB*], UNETR [56.5 *s*, 3010 *MB*], MSSegDiff [9 *min* 41.8 *s*, 2816 *MB*], LST-AI [3 *min* 36.4 *s*, 10328 *MB*]. These tests were conducted on a laptop with an Intel Core i7-8550U CPU and an externally connected NVIDIA GeForce RTX 4090 GPU (24 *GB* VRAM).

In the test on MSSEG 2016, SWIMSS24, the SwinUNETR configuration with patch size of 96 and feature size of 24, obtained a mean DSC of 70.36% (95% CI: 65.22-75.50), nDSC of 70.28% (95% CI: 64.96-75.59), and F1 score of 70.99% (95% CI: 66.11-75.88). For SWIMSS12, the mean DSC was 70.22% (95% CI: 65.71-74.73), the nDSC was 68.99% (95% CI: 63.39-74.60), and the F1 score was 68.41% (95% CI: 62.84-73.98) On the entire pediMS, the achieved mean DSC for SWIMSS24 was 63.10% (95% CI: 58.83-67.38), nDSC was 58.35% (95% CI: 53.56-63.15) and F1 score was 72.38% (95% CI: 69.62-75.15). On the other hand, SWIMSS12 achieved a mean DSC 64.13% (95% CI: 60.31-67.94), nDSC was 59.77% (95% CI: 55.08-64.46) and F1 score was 73.33% (95% CI: 70.68-75.97). Results of the few-shot learning for SWIMSS24 and SWIMSS12, using five randomly selected patients as test set and up to four for training, are reported in Table 3 and Table 4, respectively. Compared to the 0-shot test, SWIMSS24 gained about 0.5% mean DSC and F1 scores when finetuned on four patients.

## Discussion

This study leveraged the extensive data from a multicenter cohort, the SMSC, to develop and validate a robust model for WM lesion segmentation in MS. The weights of SWIMSS24 and SWIMSS12 have been made publicly available to facilitate research in the field, enabling their use for both direct deployment and fine-tuning.

Several architectures and configurations were evaluated, allowing for a more complete comparison on the same MRI data. The evaluation demonstrated that SWIMSS24 achieved the highest DSC, nDSC, and F1 score among the evaluated models, followed by SWIMSS12. In addition, the results on nDSC suggest that SWIMSS (and, partially, UNETR) better captures the volumetric extent of lesions, irrespective of their volume.

An important aspect of this work is the evaluation of multiple state-of-the-art approaches under the same preprocessing and testing framework. While previous comparisons between architectures are often performed across independent studies, differences in data composition and annotation protocols can introduce substantial bias. By training and evaluating UNETR, SWIMSS12, SWIMSS24, and MSSegDiff on the same dataset, this study provides a more reliable comparison of their relative capabilities. The consistently higher performance of SWIMSS24 and SWIMSS12 across both DSC and nDSC metrics suggests that transformer-based architectures can effectively exploit the spatial characteristics of MS lesions when sufficient training data are available. Conversely, the lower performance of LST-AI points to the challenges associated with deploying models trained on different datasets, as domain shifts caused by scanner characteristics, acquisition protocols, preprocessing methods, and population differences may considerably affect segmentation accuracy (24). For example, LST-AI achieved a DSC of 54% on pediMS (20), which represents an age-related domain shift, where SWIMSS gained 10% more DSC. SWIMSS's detection performance is also better with a balanced sensitivity and specificity.

Beyond improving benchmark performance, the availability of the trained model represents an important step toward reproducible research and broader adoption of automated WM lesion segmentation. The released weights allow researchers to directly apply the model to new datasets or perform transfer learning using local imaging cohorts (25). This approach may be particularly valuable for institutions with limited annotated data, where training a high-performing segmentation model from scratch is often impractical. Moreover, the possibility of fine-tuning the model provides a pathway toward adapting automated segmentation pipelines to different MR systems and acquisition protocols while maintaining the benefits of large-scale pretraining, as shown in the few-shot learning test.

UNETR, SWIMSS12 and SWIMSS24 required at most 6.0 GB of GPU memory, versus 10.3 GB for MSSegDiff and 22.6 GB for LST-AI, so high-end GPUs are not needed. Although SWIMSS12 had a slightly lower DSC than SWIMSS24 (80.84 vs 81.37), it achieved a higher DSC on pediMS (64.13% vs 63.10%), which combines a large age-related domain shift with site effects, and consumes one-third of GPU memory (2.1 GB). SWIMSS12 may therefore be a practical alternative in clinical settings with entry-level GPUs. Furthermore, CPU-only inference with these three models took less than 2 minutes per case on a low-power laptop processor using 2.1–3.0 GB of RAM, suggesting that CPU-based deployment in hospitals may be feasible.

An additional advantage of our approach is its generalizability to out-of-domain test data. The results obtained were comparable to in-domain test performance reported in the literature (19,20,26), suggesting that the preprocessing steps and the large amount of data used for training may have mitigated effects of domain shift.

Despite these strengths, several limitations should be considered. First, although the cohort includes data from multiple centers, all participating sites were located within Switzerland. Differences in scanner vendors, acquisition parameters, and patient demographics may still influence model performance. Second, the lesion annotations used for training were generated through an automated initial segmentation followed by expert correction, and residual annotation variability may remain, particularly for small lesions or lesions located in challenging anatomical regions. Third, this study focused on WM lesion segmentation, and further work is required to evaluate performance on cortical and spinal cord lesions, which represent additional relevant manifestations of MS pathology. Fourth, due to hardware limitations, we reduced the number of validation sampling timesteps from 50 to 5 when training MSSegDiff. This might have resulted in suboptimal model selection and slightly lower segmentation performance.

Future research should investigate domain adaptation strategies, uncertainty estimation (27), and integration of longitudinal imaging information to further improve robustness and clinical applicability. Additionally, prospective evaluation within clinical workflows will be necessary to determine whether automated segmentation can effectively reduce radiological workload and support quantitative biomarkers for treatment response (2). Overall, SWIMSS provides a scalable foundation for automated WM lesion segmentation in MS and demonstrates the potential of large multicenter datasets combined with modern transformer architectures to advance neuroimaging analysis.

### Code and Model Availability

The weights of SWIMSS, the trained SwinUNETR model (patch size of 96), will be made public (upon article publication) on Huggingface for use strictly under a non-commercial license. The code, which can be used to run inference or to fine-tune the model with additional user data, is available at (*https://osf.io/4mh97/overview?view_only=fba677d112f449fda2b32bdbc43e362a*).

# Tables

| | UNETR (96) | MSSegDiff (96) | SWIMSS12 (SwinUNETR, 96) | SWIMSS24 (SwinUNETR, 96) | LST-AI |
|---|---|---|---|---|---|
| **DSC** | 79.06 | 75.70 | 80.84 | **81.37** | 63.19 |
| **DSC 95% CI** | 78.33 – 79.79 | 74.97 – 76.43 | 80.11–81.57 | **80.63–82.11** | 62.08 – 64.30 |
| **nDSC** | 83.05 | 78.92 | 84.36 | **84.88** | 75.42 |
| **nDSC 95% CI** | 82.35 – 83.75 | 78.10 – 79.73 | 83.66–85.06 | **84.19–85.56** | 74.65 – 76.20 |
| **F1** | 78.99 | 76.13 | 80.12 | **81.14** | 68.33 |
| **F1 95% CI** | 78.30 – 79.68 | 75.34 – 76.93 | 79.41–80.82 | **80.44–81.85** | 67.41 – 69.26 |

Table 1. Test results of the four implemented architectures: UNETR, MSSegDiff, SwinUNETR, and LST-AI. Except for LST-AI, which was deployed at inference as suggested by the authors, the other three networks were trained and evaluated on our data, with different input patch size (specified in brackets). Reported metrics are mean DSC and nDSC (%) with 95% confidence interval. Best results per row are highlighted in bold.

| | SWIMSS24 | SWIMSS12 | UNETR (96) | MSSegDiff (96) | LST-AI |
|---|---|---|---|---|---|
| **DSC Mean (A - B)** | / | 0.53 | 2.24 | 5.67 | 18.18 |
| **DSC 95% CI** | / | 0.35-0.72 | 1.16-3.28 | 5.27-6.06 | 17.23-19.18 |
| ***p*-value FDR-adjusted** | | < 0.001 | < 0.001 | < 0.001 | < 0.001 |

Table 2. Statistical test results considering SWIMSS (patch size of 96) as reference model A and every other model configuration as model B. We reported the mean paired differences for the DSC, with their respective 95% confidence intervals.

| | k=0 | k=1 | k=2 | k=3 | k=4 |
|---|---|---|---|---|---|
| **DSC** | 63.70 | 64.02 | 64.17 | 64.16 | 64.21 |
| **DSC 95% CI** | 58.11–69.29 | 58.46–69.58 | 58.62–69.72 | 58.60–69.72 | 58.65–69.77 |
| **nDSC** | 59.88 | 60.33 | 60.56 | 60.55 | 60.63 |
| **nDSC 95% CI** | 54.00–65.76 | 54.49–66.17 | 54.74–66.38 | 54.73–66.37 | 54.81–66.45 |
| **F1** | 75.61 | 75.65 | 76.20 | 76.28 | 76.15 |
| **F1 95% CI** | 72.82–78.40 | 72.93–78.37 | 73.54–78.86 | 73.61–78.95 | 73.48–78.82 |

Table 3. Few shot learning results on the pediMS dataset. We show test results on the same 5 patients, when SWIMSS24 is used without finetuning (k=0) and with finetuning on up to 4 patients. The reported metrics are DSC, nDSC and F1 detection scores.

| | k=0 | k=1 | k=2 | k=3 | k=4 |
|---|---|---|---|---|---|
| **DSC** | 64.86 | 65.04 | 65.17 | 65.22 | 65.20 |
| **DSC 95% CI** | 59.88-69.84 | 60.15- 69.93 | 61.39-69.16 | 60.33-70.01 | 60.41-70.03 |
| **nDSC** | 60.91 | 61.37 | 61.71 | 61.79 | 61.79 |
| **nDSC 95% CI** | 55.24-66.58 | 55.62- 67.12 | 55.98- 67.44 | 56.06-67.52 | 56.06-67.52 |
| **F1** | 76.95 | 76.14 | 76.30 | 76.51 | 76.46 |
| **F1 95% CI** | 74.23-79.67 | 73.33-78.95 | 73.60-79.00 | 73.63-79.39 | 73.60-79.32 |

Table 4. Few shot learning results on the pediMS dataset. We show test results on the same 5 patients, when SWIMSS12 is used without finetuning (k=0) and with finetuning on up to 4 patients. The reported metrics are DSC, nDSC and F1 detection scores.

## Disclosures

Federico Spagnolo has nothing to disclose.

Po-Jui Lu has nothing to disclose.

Mario Ocampo-Pineda has nothing to disclose.

Lester Melie-Garcia has nothing to disclose.

Esther Ruberte has nothing to disclose.

Riccardo Galbusera is supported by a Postdoc Mobility Grant from the Swiss National Science Foundation (SNSF).

Matthias Weigel received research funding from Biogen in the past.

Alessandro Cagol received speaker honoraria from Merck, Novartis, and Roche.

Johanna Maria Lieb has nothing to disclose.

Nicolin Hainc has nothing to disclose.

Franca Wagner has nothing to disclose.

Piotr Radojewski has nothing to disclose.

Felix Tobias Kurz has nothing to disclose.

Vincent Dunet has nothing to disclose.

Emanuele Pravatà has nothing to disclose.

Olaf Chan-Hi Kim has nothing to disclose.

Patrick Roth received honoraria for lectures or advisory board participation from Alexion, Bristol-Myers Squibb, Boehringer Ingelheim, Debiopharm, Galapagos, Merck Sharp and Dohme, Laminar, Midatech Pharma, Novocure, QED, Roche, Sanofi and Servier and research support from Merck Sharp and Dohme and TME Pharma.

Marina Herwerth served on scientific advisory boards and/or received speaker's honoraria, travel support from Biogen, Merck Serono, Alexion, Roche and Horizon Therapeutics (Amgen), unrelated to this study. Her institution received a research grant from Roche. She was supported by the Swiss National Science Foundation (SNSF, PZ00'3_216616/1) and by the Olga-Mayenfisch-Foundation.

Sebastian Finkener has nothing to disclose.

Robert Hoepner received speaker/advisor honorary from Merck, Novartis, Roche, Biogen, Alexion, Sanofi, Janssen, Bristol-Myers Squibb, Teva/Mepha and Almirall. He received research support within the last 5 years from Roche, Merck, Sanofi, Biogen, Chiesi, and Bristol-Myers Squibb. He also received research grants from the Swiss MS Society, the SITEM Insel Support Fund and is a member of the Advisory Board of the Swiss and International MS Society. He also serves as deputy editor in chief for Journal of Central Nervous System Disease and is part of the ECTRIMS Young Investigator Committee.

Patrice H. Lalive has nothing to disclose.

Caroline Pot has nothing to disclose.

Claudio Gobbi has nothing to disclose.

Stefanie Müller received honoraria for travel, honoraria for lectures/consulting and/or grants for studies from Almirall, Alexion, Bayer, Biogen, Bristol-Myers Squibb SA/Celgene, Genzyme, Merck-Serono, Teva, Novartis and Roche. O.C.-H. Kim reports no disclosures relevant to the manuscript.

Jens Kuhle received speaker fees, research support, travel support, and/or served on advisory boards by Swiss MS Society, Swiss National Research Foundation (320030_212534/1), University of Basel, Progressive MS Alliance, Alnylam, Bayer, Biogen, Bristol Myers Squibb, Celgene, Immunic, Merck, Neurogenesis, Novartis, Octave Bioscience, Quanterix, Roche, Sanofi, Stata DX.

Vincent Andrearczyk has nothing to disclose.

Adrien Depeursinge has nothing to disclose.

Cristina Granziera as the employer of the University Hospital Basel (USB) and Research Center for Clinical Neuroimmunology and Neuroscience Basel (RC2NB), has received the following fees which were used exclusively for research support: (i) advisory boards, and consultancy fees from GeNeuro, Actelion, Novartis, Genzyme-Sanofi, Biogen, Hoffmann La Roche and Siemens; (ii) speaker fees from Biogen, Hoffmann La Roche, Teva, Novartis, Merck, Janssen Pharmaceuticals and Genzyme-Sanofi; (iii) research grants: Biogen, Genzyme Sanofi, Hoffmann La Roche, Novartis, GeNeuro.

## Figures

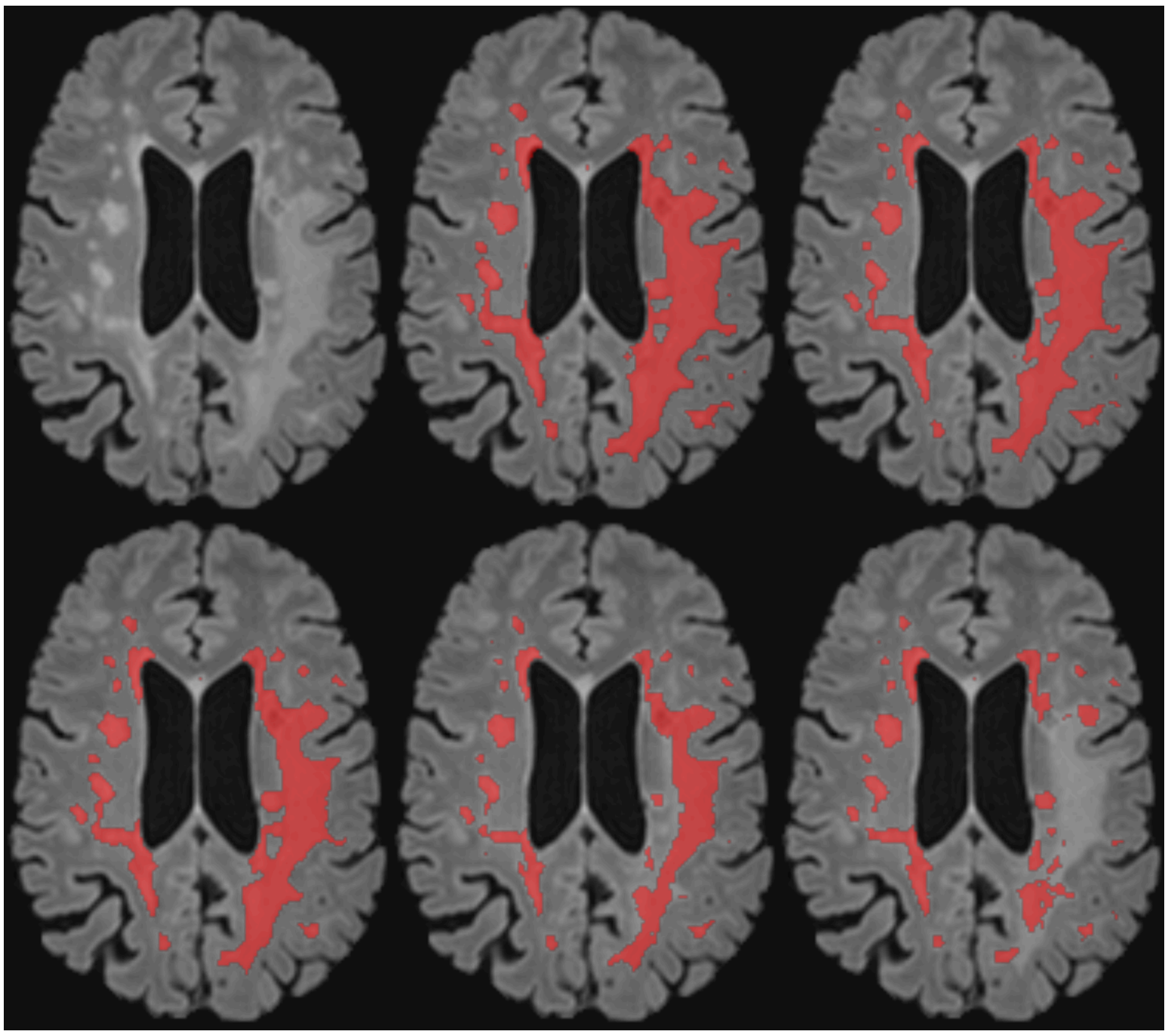

Figure 1. Predicted lesion masks for the test case with the highest lesion volume. From top to bottom, left to right: original FLAIR, ground truth mask, SWIMSS24, UNETR, MSSegDiff, LST-AI.

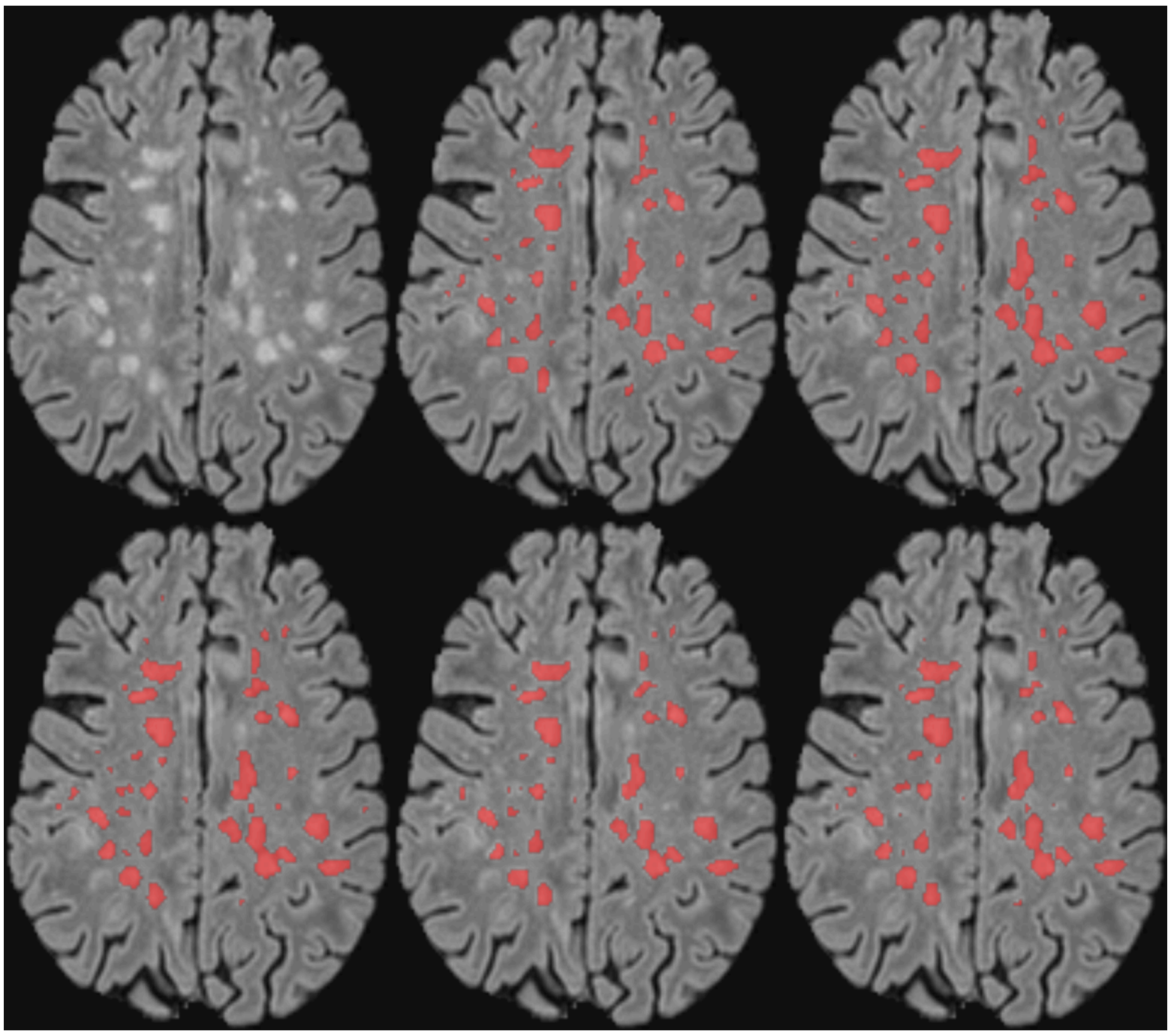

Figure 2. Predicted WM lesion masks for the test case with the highest lesion count. From top to bottom, left to right: original FLAIR, ground truth mask, SWIMSS24, UNETR, MSSegDiff, LST-AI.